# Student understanding of time in special relativity: simultaneity and reference frames

Rachel E. Scherr, Peter S. Shaffer, and Stamatis Vokos
Department of Physics, University of Washington, Seattle, WA

This article reports on an investigation of student understanding of the concept of time in special relativity. A series of research tasks are discussed that illustrate, step-by-step, how student reasoning of fundamental concepts of relativity was probed. The results indicate that after standard instruction students at all academic levels have serious difficulties with the relativity of simultaneity and with the role of observers in inertial reference frames. Evidence is presented that suggests many students construct a conceptual framework in which the ideas of absolute simultaneity and the relativity of simultaneity harmoniously co-exist.

## I. INTRODUCTION

There is growing national interest in increasing the exposure of students in introductory courses to modern physics topics, such as relativity. Proponents of enlarging the scope of the curriculum argue that in the beginning of the 21st century the content of introductory classes should reflect some of the major intellectual breakthroughs of the 20th century. Others hold that the list of topics that must be covered is already too daunting. Physics education research can play a pivotal role in this debate. Whether students first encounter modern physics concepts at the introductory level or in advanced courses, it is important to identify what students can and cannot do after instruction and what steps can be taken to help deepen their understanding of the material. In addition, analyzing the ways in which students undergo the transition between understanding phenomena to which they have immediate access and understanding phenomena that lie outside their everyday experience can help us identify reasoning skills that are needed for the study of advanced topics.

Over the last five years, the Physics Education Group at the University of Washington has been investigating student understanding of key ideas in Galilean, special, and general relativistic kinematics. Extensive research has already been conducted on student understanding of non-relativistic kinematics in the laboratory frame.[1] We wanted to expand this research base to relativity in order to provide a guide for the development of instructional materials by ourselves and others.[2,3]

This article reports on an investigation of student understanding of time in special relativity. A major purpose is to identify and characterize the conceptual and reasoning difficulties that students at all levels encounter in their study of special relativity. The emphasis is on the relativity of simultaneity and the role of reference frames. We found that, after instruction, many students are unable to determine the time at which an event occurs, recognize the equivalence of observers at rest relative to one another, and apply the definition of simultaneity. We illustrate the process through which we gradually obtained a detailed picture of student thinking by the design and successive refinement of a set of research tasks.

## II. PRIOR RESEARCH

There is currently only a small body of research on student understanding of relativity, mostly in Galilean contexts (in particular, relative motion). In many cases, the research tasks used (*e.g.,* multiple-choice questions or single questions given to a small number of students) do not provide the kinds of insights that are necessary to understand the prevalent modes of student reasoning and to develop effective instructional strategies. We review salient results from relevant investigations.

*Galilean relativity*

Investigations of physics undergraduates in India have identified the belief that reference frames have limited physical extent. Studies by Panse *et al.* and Ramadas *et al.* suggest that, for many students, frames of reference have limited physical extent. Students claim that a body can "emerge" from the reference frame of an object by leaving the vicinity of the object (*e.g.,* "a ball can be thrown to 'go outside' a reference frame").[4]

An investigation by Saltiel and Malgrange has identified difficulties with relative motion among eleven-year-old-children and first- and fourth-year university students in France.[5] The three groups showed little difference in error rates to written questions. Many students tended to identify an object's motion as intrinsic, not a quantity that is measured relative to a reference frame. Students tended to make a distinction between "real" motion, which has a dynamical cause, and "apparent" motion, which is "an optical illusion, devoid of any physical reality."

*Special relativity*

Villani and Pacca have demonstrated that university students' reasoning in relativistic contexts is similar to that observed by Saltiel and Malgrange in Galilean contexts.[6] A case study by Hewson with a physics graduate student illustrated the importance of "metaphysical beliefs" (*e.g.,* time is absolute) to his understanding of special relativity.[7] The student in the study classified certain relativistic effects (including length contraction) as distortions of perception. Posner *et al.* report similar results in interviews with introductory students and their instructors.[8]

O'Brien-Pride has conducted interviews and administered early versions of some of the research tasks described here in which university students appear to believe that the order of events depends on observer location.[9] Her preliminary results provided impetus for the investigation detailed in this paper.

## III. FOCUS OF THE RESEARCH

A major goal of the investigation was to determine the extent to which students, after instruction, are able to apply basic ideas of special relativity to simple physical situations. The context is one spatial dimension. The concepts probed are summarized below.

The construct of a *reference frame* is at the heart of relative motion. Most courses in special relativity begin with a discussion of a reference frame as a system of observers (or devices) by which the positions and times of events are determined. Understanding the concept of a reference frame forms the foundation for understanding any topic in special relativity. It provides the basis for the determination of all kinematical (and other physical) quantities and serves as the framework for relating measurements made by different observers. The concept of a reference frame presupposes an understanding of more basic measurement procedures. For clarity in the discussion that follows, we review the basic operational definitions associated with reference frames, *i.e.,* the determination of the position and time of an event and the conditions under which two events are treated as simultaneous.[10]

An *event* in special relativity is associated with a *single* location in space and a *single* instant in time. The *position of an event* is defined to be the coordinate label on a rigid ruler at the location of the event. The ruler is envisioned to extend indefinitely from some chosen origin.[11] The *time of an event* is most naturally defined as the reading on a clock located at the event's position at the instant at which the event occurs. The rulers and clocks used by any observer are at rest relative to the observer.

All observers in special relativity are assumed to be "intelligent observers" who use synchronized clocks. To determine the time of a distant event, an observer corrects for the travel time of a signal originating at the event.[12,13] Inertial observers at rest relative to one another determine the same positions and times for events (and hence the same relative ordering of events). Such observers are said to be *in the same reference frame*.[14]

Events are defined to be *simultaneous* in a given frame if their corresponding time readings are identical, according to the definition of the time of an event discussed above. A judgment of the simultaneity of widely separated events necessarily includes an appropriate definition of the time of a distant event.[15] The invariance of the speed of light has an inevitable and unsettling implication. Two events at different locations that occur at the same time in a given frame are not simultaneous in any other frame.

The relativity of simultaneity is among the key results of special relativity and one that is particularly difficult to grasp, as evidenced by the numerous "paradoxes" that arise from it. It is not reasonable to expect that students (even those facile with mathematical formalism) will master all the intricacies of the counterintuitive results that follow from the operational definition of simultaneity. In this study, we considered a meaningful understanding of relativistic simultaneity to include the ability to identify relevant events, to determine the time at which an event occurs (by correcting for signal travel time), and to recognize that the time interval between two events is not invariant but depends on the reference frame. Other aspects of student ideas about time and reference frames are not discussed (*e.g.*, synchronization of clocks in relative motion, spatial measurements, differences between inertial and non-inertial frames).

## IV. OVERVIEW OF THE RESEARCH

This investigation was conducted over the last five years at the University of Washington and at two other large research universities. Fourteen instructors at the University of Washington and one faculty member at each of the other universities have cooperated with the Physics Education Group in this study.

### A. Student populations

Most of the research was conducted at the University of Washington in courses that include special relativity. The study has involved about 800 students from about 30 sections of various courses. The populations include: non-physics students (in the descriptive course for non-majors students); introductory students (in the introductory calculus-based honors course and in the sophomore-level course on modern physics); students in the junior-level courses on electricity and magnetism and relativity and gravitation; and students in our upper-division course for prospective high school physics teachers.[16] We also present results from physics graduate students at the University of Washington who participated in interviews and others who were given a written question on a graduate qualifying examination. In addition, the investigation includes students in the honors section of the calculus-based course at one of the other research universities and advanced undergraduate students from the other collaborating university. We found that student performance from all three universities was similar. The results, therefore, have been combined.[17]

### B. Research methods

The research was conducted through the analysis of student responses to written questions and the analysis of interviews with individual students. The written questions were posed on course examinations and on ungraded problems given during class.[18] In some classes the questions were administered before instruction on the relevant concepts; in other classes, the questions were administered after instruction. Except where otherwise noted, no class was given a question twice.

The one-hour interviews were conducted with volunteers, who were primarily first-year physics graduate students. Several advanced undergraduate students and a few advanced graduate students also participated. In the interviews, we were able to probe student thinking in greater depth than is possible with short written questions. The interviews were audiotaped or videotaped and transcribed for later analysis.

### C. Research tasks

To obtain an increasingly deeper understanding of how students apply the concepts of simultaneity and reference frame, we used variants of three questions: the *Spacecraft* question (four versions), the *Explosions* question, and the *Seismologist* question. All involve two observers with a given relative motion. Students are told the time ordering of the events for one observer and asked about the time ordering of the events for the second observer.[19] These questions and their solutions are described below.

#### 1. *Spacecraft* question

Results from four versions of the *Spacecraft* question are discussed in this paper. All involve two volcanoes, Mt. Rainier and Mt. Hood, that erupt simultaneously according to an observer at rest on the ground, midway between the volcanoes.[20] A spacecraft moves at a given relativistic velocity from Mt. Rainier to Mt. Hood.[21] Students are asked questions that probe their beliefs about the order of the eruptions in the moving frame.

A correct answer to all versions can be obtained by qualitative or quantitative reasoning or from a spacetime diagram. The following is an example of a qualitative argument that we would have accepted as correct.[22] In the spacecraft frame, light from the two eruptions moves outward at the speed of light in spherical wavefronts from two points that are stationary. In that frame, the observer on the ground, who receives both signals simultaneously, is moving backward (*i.e.*, in the direction of an arrow pointing from the front of the spacecraft toward the rear). According to the spacecraft observer, the ground-based observer is closer to the center of the signal from Mt. Rainier at the instant that observer receives both signals. The spacecraft observer therefore concludes that Mt. Hood erupted first since its signal travels farther in order to reach the ground-based observer at the same time as the signal from Mt. Rainier.

A correct answer can also be obtained using the Lorentz transformation for time: $dt' = g(dt - vdx/c^2)$, $g = (1 - v^2/c^2)^{-1/2}$. In this context, $dt' = t'_H - t'_R$ and $dt = t_H - t_R$ are the elapsed times between the eruptions at Hood and Rainier in the spacecraft frame and the ground frame respectively, $v$ is the velocity of the spacecraft relative to the ground, and $dx = x_H - x_R$ is the spatial coordinate separation between the eruptions in the ground frame.[23] Taking the positive direction to be directed from Rainier to Hood, then $v > 0$ and $dx > 0$. Since $dt = 0$ (simultaneous eruptions in the ground frame), then $dt' < 0$.

#### 2. *Explosions* question

The *Explosions* question is the converse of the *Spacecraft* question. Students are told that two events occur at different times in a given frame and are asked if there is another frame in which the events are simultaneous.

In the *Explosions* question, an explosion occurs at each end of a landing strip with proper length of 3000 m. In the frame of an engineer at rest on the strip, the explosion at the right end occurs a time $c\ dt = 1200$ m after the explosion on the left end. (In some variations, students were given a time of $dt = (1200\ \text{m})/c = 4\ \text{ns}$. The conversion seemed to present no difficulty.) Students are asked whether there is a frame in which the explosions are simultaneous, and if so, to determine the relative velocity of that frame.

A correct answer can be found through use of the Lorentz transformations.[24] The spatial separation between the explosions ($dx$) is 3000 m and the time separation ($c\ dt$) is 1200 m. Thus the time duration between the explosions $(c\ dt')$ is zero in a frame that moves from left to right with speed $0.4c$.

#### 3. *Seismologist* question

The *Seismologist* question probes student understanding of reference frames and simultaneity within a single reference frame. The context is similar to that of the *Spacecraft* question: two volcanoes, Mt. Rainier and Mt. Hood, suddenly erupt and a seismologist at rest midway between them sees the eruptions at the same instant. The *Seismologist* question differs from the *Spacecraft* question in that the second observer (the "assistant") is not moving, but remains at rest relative to the ground at the base of Mt. Rainier. Students are asked whether Mt. Rainier erupts before, after, or at the same instant as Mt. Hood in the reference frame of the assistant.

To answer the *Seismologist* question correctly, students must be able to apply the definition of simultaneity and understand the role of a reference frame in establishing a common time coordinate for observers at rest relative to one another. Since the seismologist and the assistant are in the same reference frame, they obtain the same answer for the order of the eruptions. Since the seismologist is equidistant from the mountains, the signal travel times are the same. Therefore, the eruptions occurred at the same time in the frame of the seismologist and the assistant.

*Commentary*

At each stage of our study, we tried to determine whether student responses truly reflected their understanding of the material. For instance, we wanted to determine the extent to which specific difficulties are linguistic or conceptual and the extent to which mistaken beliefs are easily addressed or deeply held. To this end, we continually refined the research tasks. Results from earlier tasks guided us in designing new questions that would enable us to probe student thinking more thoroughly.

## V. PRELIMINARY INVESTIGATION OF STUDENT UNDERSTANDING OF THE CONCEPTS OF SIMULTANTEITY AND REFERENCE FRAMES

Our preliminary investigation of student understanding of special relativity was based on two versions of the *Spacecraft* question: an *undirected* and a *directed* version. The two versions and the results from each are described below.

### A. Failure to recognize spontaneously that simultaneity is relative

We refer to the first version of the *Spacecraft* question as *undirected*. We were interested in finding out whether or not students would recognize, without prompting, that simultaneity is relative and, if not, the degree of prompting that is necessary for them to apply the relativity of simultaneity.

*Spacecraft Question: Undirected version*

The students were asked to draw spacetime diagrams for both the ground and spacecraft frames. They were told to show the volcanoes, the spacecraft, and the eruption events. They were not asked explicitly whether the eruption events are simultaneous in the spacecraft frame. Rather, we inferred their ideas indirectly from their diagrams.

We administered this *undirected* version as an interview task to 7 graduate students and later to 20 advanced undergraduate students enrolled in a course in special and general relativity. All the graduate students had had undergraduate instruction in special relativity and were studying relativistic kinematics, dynamics, and electromagnetism in their graduate-level electricity and magnetism course at the time of the interviews. The undergraduates had completed instruction on the relativity of simultaneity. All had worked with spacetime diagrams in their current or previous courses.

All the graduate students correctly drew the worldlines of each object in the spacetime diagram for each frame. However, only one recognized spontaneously the relativity of simultaneity in this context. All the others indicated on their spacetime diagrams and in their verbal explanations that the two eruptions had identical vertical (time) coordinates in the ground frame *and* identical time coordinates in the spacecraft frame. [See Fig. 1 for examples of correct and incorrect spacetime diagrams.] The results from the undergraduate students were similar. All drew spacetime diagrams that included the correct features except that about 85% denoted the eruption events as simultaneous in both frames. Some of the difficulty may have been related to a lack of facility in relating spacetime diagrams for two frames. Nonetheless, the majority of the students failed to recognize spontaneously that simultaneity is relative and to draw their diagrams appropriately.

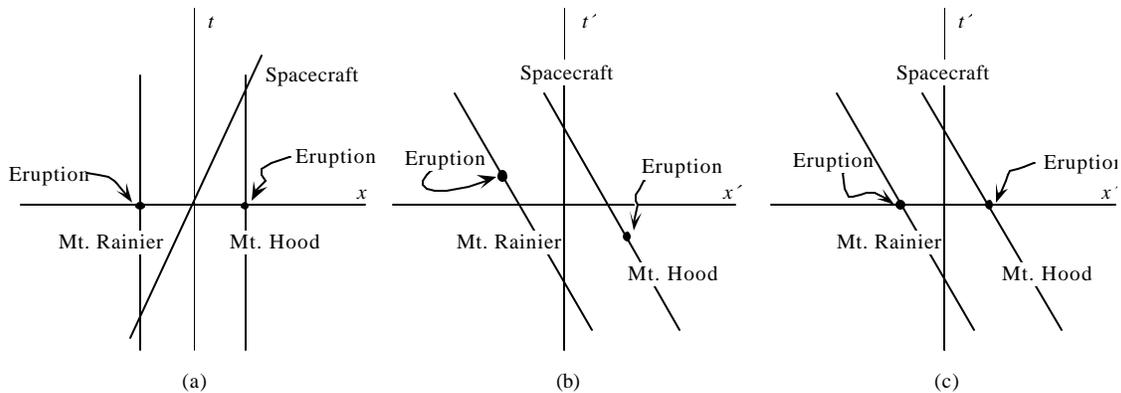

Figure 1: Spacetime diagrams for the first *(undirected)* version of the *Spacecraft* question. (a) Correct diagram for the ground frame. (b) Correct diagram for the spacecraft frame. (c) Typical incorrect diagram for the spacecraft frame drawn by students.

The relativity of simultaneity is arguably the central result of relativistic kinematics and a key consequence of the Lorentz transformations. The fact that advanced students do not apply this idea spontaneously is a matter of concern. On the other hand, the fact that the time order of events is not the same in all frames is among the most counterintuitive ideas in special relativity. We wondered whether students might apply the simultaneity of relativity if prompted explicitly to do so.

**B. Failure to apply spontaneously the formalism of a reference frame in determining the time of an event**

We decided to develop a new version of the *Spacecraft* question that would be more directive than the first. This version was written in collaboration with the course instructor. The terminology was identical to that used in class.

*Spacecraft Question: Directed version*

In the *directed* version of the *Spacecraft* question, students are asked explicitly whether, in the reference frame of the spacecraft, Mt. Rainier erupts before, after, or at the same time as Mt. Hood. They are also asked to find the time between the eruptions.

We administered this *directed* version of the *Spacecraft* question to 49 students on an examination in an introductory honors calculus-based physics class. The students had completed the study of the relevant material. All students answered that the eruptions are not simultaneous in the spacecraft frame. (Very few students had done so on the undirected version.) However, only about 45% gave the correct time ordering (Hood erupts first in the spacecraft frame) with correct reasoning.[25] About 25% of the students gave the correct time order of events but used incomplete reasoning to support their answers. About 25% of the class gave the reverse time ordering.

Essentially all of the students who gave a correct answer with incomplete reasoning answered in a similar way. The responses below are typical.

> "Mt. Hood erupted first because the spacecraft is moving towards it, so the wavefront of the eruption of Mt. Hood will reach the craft first." (introductory student)

"Hood first, because the spacecraft will encounter those wavefronts first." (introductory student)

We categorize the reasoning given by these students as incomplete since these students describe only the order in which the signals from the distant events reach the spacecraft. They make no explicit mention of the relative velocity or the relativity of simultaneity. The sequence in which the signals are received does not provide enough information to determine the time ordering of the eruption events. Different choices of observer location result in different reception orders and some students might have obtained the correct answer by a fortuitous choice of observer location (*e.g.*, half-way between the two volcanoes). We started to suspect the presence of incorrect ideas when we realized that most of the remaining students (about 25%) seemed to have made a different assumption about the spacecraft location. These students obtained the opposite answer for the time ordering of the events but gave explanations similar to those illustrated above. The following response was typical.

"The observer will witness Mt. Rainier erupting first because they are directly over Rainier when the explosion happens, so the light from the explosion has less distance to travel than the light from Mt. Hood's explosion." (introductory student)

Several students regarded both the velocity and position of the spacecraft as important in the time ordering of the eruption events. However, like the students above, they focused on the reception of the light signals by the observer. They did not treat relative motion as the determining feature of a reference frame but as a factor that complicates the calculation of the time at which the observer receives the signals. One student claimed necessary information was missing from the problem statement.

"It would depend on where the spacecraft was when the first explosion occurs. If it is close enough to Hood that the distance between the ship and Hood plus the distance the ship travels while the light is en route, then it sees Hood explode first." (introductory student)

The students quoted above all failed to treat the spacecraft observer as representative of a class of observers, all moving with the same velocity. They seemed to interpret the time of an event as an observer- (not frame-) dependent quantity. These results suggest that students, on their own, fail to apply the formalism of reference frames (*i.e.*, a system of clocks and rods) in defining the time of an event.

*Commentary*

The versions of the *Spacecraft* question used in the preliminary investigation are similar to many end-of-chapter questions on relativistic simultaneity. Students are told that two events are simultaneous in one frame *(S)* and asked about the time order of the events in another frame *(S′)* that moves relative to the first with a given velocity.

In the context of the Spacecraft question, we found that many students fail to apply spontaneously the relativity of simultaneity. When prompted to think explicitly about the order of the events, essentially all students state that the events are not simultaneous in the spacecraft frame. However, most reason incorrectly. They tend to focus on the relative *position* of the spacecraft and volcanoes and fail to recognize that the relative *velocity* determines the time order of the eruptions in the spacecraft frame.

The results from the undirected and directed versions of the *Spacecraft* question suggested the presence of serious conceptual and reasoning difficulties with basic concepts in special relativity. We considered, however, the possibility that student responses did not reflect what students actually thought. We needed to probe more deeply into the nature of their conceptions of time, simultaneity, and reference frames.

## VI. DETAILED INVESTIGATION OF STUDENT UNDERSTANDING OF THE CONCEPTS OF TIME, SIMULTANEITY, AND REFERENCE FRAMES

This section is divided into three inter-related and inter-dependent parts. Part A deals primarily with a prevalent and persistent student interpretation of simultaneity that is observer-dependent. Many students believe that the time order of distant events is determined by the time order in which signals from the events are perceived by an observer. Part B presents evidence that many students have a deeply held underlying belief that simultaneity is absolute. Part B also describes how students often attempt to reconcile these two contradictory beliefs with each other and with what they have been taught about the relativity of simultaneity. Student interpretations and beliefs about simultaneity described in Parts A and B have direct implications on student understanding of the role of an observer in a given frame. Part C is devoted to an exploration of student beliefs about the concept of a reference frame.

### A. Belief that events are simultaneous if an observer receives signals from the events at the same instant

We decided that the *Spacecraft* question would be a better probe of student thinking about simultaneity if two changes were made: (1) specifying not only the velocity but also a location for the moving observer and (2) rewording the question to describe the eruptions as spacetime events. By choosing the observer location appropriately, we sought to distinguish between students who obtained a correct answer for correct reasons and students who

---

> Mt Rainier and Mt. Hood, which are 300 km apart in their rest frame, suddenly erupt at the same time in the reference frame of a seismologist at rest in a laboratory midway between the volcanoes. A fast spacecraft flying with constant speed $v = 0.8c$ from Rainier toward Hood is directly over Mt. Rainier when it erupts.
>
> Let Event 1 be "Mt Rainier erupts," and Event 2 be "Mt. Hood erupts."
>
> In the reference frame of the spacecraft, does Event 1 occur *before, after,* or *at the same time as* Event 2? Explain your reasoning.

Figure 2: *Location-specific* version of the *Spacecraft* question.

thought the observer's location affects the time ordering of the events. By describing the eruptions as events, we tried to make clear to students that the time interval of interest is not that between the reception of the light signals by the moving observer, but rather that between the emission of the signals by the volcanoes.

*Spacecraft question: Location-specific version*

In the third *(location-specific)* version of the *Spacecraft* question, students are told that the spacecraft, which is flying from Mt. Rainier to Mt. Hood, is over Mt. Rainier at the instant Mt. Rainier erupts. The eruption events, which are simultaneous in the ground frame, are explicitly labeled as Event 1 (Mt. Rainier erupts) and Event 2 (Mt. Hood erupts). Students are to determine whether, in the reference frame of the spacecraft, Event 1 occurs before, after, or at the same time as Event 2. [See Fig. 2.]

Table I summarizes the results of the location-specific *Spacecraft* question, which was administered as an ungraded written question to non-physics students, introductory students, and advanced undergraduate students. The question was also given as an interview task to advanced undergraduates and physics graduate students. In most cases, the question was posed after lecture instruction on the relativity of simultaneity. As can be seen from the table, prior instruction had little effect on student performance.[26]

Table I. Results from the third *(location-specific)* version of the *Spacecraft* question. Students are told the eruptions are simultaneous for one observer and are given the location of an observer in a second frame moving relative to the first. They are asked for the order of the eruptions in the frame of the second observer. [Percentages have been rounded to the nearest 5%.]

|  | Written question |  |  |  |  |  | Interview task |
|---|---|---|---|---|---|---|---|
|  | Non-physics students | | Introductory students (Honors calculus-based physics) | | Advanced undergraduates | | Advanced undergraduates and graduate students |
|  | Before instruction[†] (N = 23) | After instruction (N = 16) | Before instruction (N = 67) | After instruction (N = 73) | Before instruction (N = 20) | After instruction[‡] (N = 93) | (N = 11) |
|  | % (N) | % (N) | % (N) | % (N) | % (N) | % (N) | % (N) |
| Correct answers (Hood erupts first) with correct reasoning *or* incomplete reasoning [*] | 15% (3) | 15% (2) | 5% (3) | 10% (8) | 15% (3) | 25% (24) | 25% (3) |
| Simultaneous eruptions (reasoning consistent with being based on absolute simultaneity) | 45% (10) | 30% (5) | 20% (12) | 5% (5) | 25% (5) | 20% (19) | 0 (0) |
| Rainier erupts first (reasoning consistent with being based on the times at which signals are received by the observer) | 35% (8) | 45% (7) | 70% (46) | 75% (55) | 45% (9) | 40% (39) | 55% (6) |
| Other (*e.g.,* student stated not enough information given) | 10% (2) | 15% (2) | 10% (6) | 5% (5) | 15% (3) | 10% (11) | 20% (2) |

[*] Some students gave a correct answer with reasoning that was incomplete, but not incorrect. Although it was not possible to tell whether they were correct in their reasoning, in this article the responses are treated as correct.

[†] These students had received research-based instruction on reference frames in Galilean relativity in which they had developed a definition for the time of an event. We have evidence that this special instruction may have been responsible for the low percentage of students answering that Rainier erupts first. The students had had no instruction in special relativity.

[‡] One of the five classes of advanced undergraduates who were given the location-specific version of the *Spacecraft* question had previously been given a variant of the question as a pretest. The results were comparable to the other four classes.

In every student group, fewer than 25% of the students gave a correct response (ignoring reasoning). Of those who gave the correct order, most used complete reasoning. Only a few used incomplete reasoning similar to that in the preliminary investigation (*e.g.,* "Hood erupts first since the spacecraft is flying towards Hood.").

The majority of students in every student group answered incorrectly. Analysis of student responses revealed two related modes of incorrect reasoning.

### 1. *Tendency to associate the time of an event with the time at which an observer receives a signal from the event*

Despite having been told explicitly that the events of interest are Event 1 (Mt. Rainier erupts) and Event 2 (Mt. Hood erupts), many students attributed the time of each eruption to the time at which a given observer *sees* the eruption. Both the advanced students in the interviews and the introductory students on the written problems made this error. The following statements are typical.

> "The spacecraft is near Rainier, so he gets the signal about the same time Rainier erupts. So the spacecraft pilot would say Rainier erupts before Hood." (graduate student)

> "Mt. Rainier erupts first because the light from Mt. Hood takes time to reach the spaceship." (introductory student)

### 2. *Tendency to regard the observer as dependent only on his or her personal sensory experiences*

The failure to distinguish between the time of an event and the time at which an observer sees that event occurring did not seem to be a superficial error but seemed to have deep roots. Many students failed to recognize that an observer is not isolated but has access to information provided by other observers in his or her frame.

> "For example if he looks at [the volcano, it] looks peaceful. There is nothing going on. So he would say it hasn't erupted. I would say, to state that something happened you have to have any evidence, and he hasn't got any evidence that something happened. So it hasn't happened." (graduate student)

*Commentary*

The third, *location-specific* version of the *Spacecraft* question provided insights into student thinking about simultaneity that the first two versions had not. The responses to the written questions and interviews show that many students very strongly associate the time of an event with the time at which an observer receives a signal from the event. Whether or not distant events are simultaneous is, therefore, judged by many students only on the basis of the time order of the received signals. The difficulties seem to be intimately tied to their ideas of reference frames.

The tendency of students to interpret simultaneity in terms of signal reception had, thus far, prevented us from determining whether or not the students recognized that the events themselves are not simultaneous in all frames. We thought, however, that the failure to use the emission events might be easy to correct with upper-level undergraduate students and graduate students. We wondered whether students would apply the relativity of simultaneity properly if, during interviews, it were pointed out to them that the reception events are not the ones to consider.

### B. Belief that simultaneity is absolute

As described earlier, we found that many students fail to apply spontaneously the idea of relativity of simultaneity after instruction. On the other hand, we also found that many students appear to believe in a type of relativity of simultaneity that we would term excessive. Students often claim that observers at different locations determine different time orderings for events based on the reception of signals from the events. We now present evidence that these apparently contradictory beliefs (that simultaneity is absolute, and that simultaneity is "excessively" relative) often coexist harmoniously. Our results suggest that many students believe that simultaneity is relative *only* in the limited sense that signals from events arrive at different observers at different times – and that fundamentally, simultaneity is absolute.

*Spacecraft question: Explicit version*

We wanted to ensure that students were not hindered by semantic misinterpretations of technical terms such as "intelligent observer" and "reference frame." In an effort to remove possible ambiguity from the task and to probe even more explicitly than before, we designed a fourth version of the *Spacecraft* question. We refer to it as the

> In this problem, all events and motions occur along a single line in space. Non-inertial effects on the surface of the Earth may be neglected.
>
> Two volcanoes, Mt. Rainier and Mt. Hood, are 300 km apart in their rest frame. Each erupts suddenly in a burst of light. A seismologist at rest in a laboratory midway between the volcanoes receives the light signals from the volcanoes at the same time. The seismologist's assistant is at rest in a lab at the base of Mt. Rainier.
>
> Define Event 1 to be "Mt. Rainier erupts," and Event 2 to be "Mt. Hood erupts."
>
> A fast spacecraft flies past Mt. Rainier toward Mt. Hood with constant velocity $v = 0.8c$ relative to the ground ($g = 5/3$). At the instant Mt. Rainier erupts, the spacecraft is directly above it and so the spacecraft pilot receives the light from Mt. Rainier instantaneously.
>
> All observers are *intelligent* observers, *i.e.,* they correct for signal travel time to determine the time of events in their reference frame. Each observer has synchronized clocks with all other observers in his or her reference frame.
>
> For each intelligent observer below, does Event 1 occur *before, after,* or *at the same time as* Event 2? Explain.
>
> - Seismologist
> - Seismologist's assistant
> - Spacecraft pilot

Figure 3: *Explicit* version of the *Spacecraft* question.

*explicit Spacecraft* version because it makes explicit the correction for signal travel time employed by intelligent observers. [See Fig. 3.]

We initially administered the question as an interview task so that we could carefully observe and correct, as necessary, the way in which students interpreted the question. We probed both qualitative and quantitative reasoning. The interviews were conducted with 2 advanced undergraduates and 5 graduate students. The question was subsequently given to 23 graduate students as part of a written qualifying examination for doctoral candidacy.

In the *explicit* version of the *Spacecraft* question, students are told that "observers are intelligent observers, *i.e.,* they correct for signal travel time in order to determine the time of events in their reference frame. Each observer has clocks that are synchronized with those of all other observers in his or her reference frame." In the course of the interview, students were reminded repeatedly to consider all observers as making corrections for signal travel time. When students used technical terms such as "reference frame," the interviewer probed their understanding of the term. If a student's interpretation differed from the conventional interpretation, the interviewer attempted to correct the student, and asked the student to reconsider his or her response in light of the accepted interpretation.

The results from the interviews and the written doctoral examination were similar. In the interviews, a correct answer was given by 1 of the 2 advanced undergraduates, and 2 of the 5 graduate students. On the written question, 7 of the 23 graduate students (30%) answered correctly. The quotes given in the discussion below are from the interviews since the interview format allowed us to probe student reasoning in greater detail than is possible in a written question.

1. ***Tendency to regard the relativity of simultaneity as an artifact of signal travel time***

During the interview, many students seemed to resist thinking about simultaneity in terms of emission, rather than reception, of the signals. As the interviews progressed, we realized that part of the difficulty was that they believed strongly that, in every reference frame, the two events occurred at the same instant. Many seemed to treat the non-simultaneity of the reception of the signals as a way of reconciling this belief with what they thought they had learned about the relativity of simultaneity.

Four of the seven advanced undergraduate and graduate students who responded to the explicit *Spacecraft* question clearly articulated the idea that the order of events in the spacecraft frame is determined by the order in which the signals from the events arrive at the spacecraft. (About 60% of the graduate students did so on the qualifying examination.) Their explanations were similar to those quoted previously in response to the other versions of the *Spacecraft* question. When reminded to consider the spacecraft observer as making corrections for the signal travel time, all of these interview subjects claimed that after making such corrections, the intelligent observer in the spacecraft would determine the eruptions to have been simultaneous.

> "Using her correction, assuming she's intelligent…I mean if she measured the effect relativistically, she would measure them happening at the same time if she subtracted the time she calculated." (graduate student)

> "If we are in relative motion we will measure different distances and so on but if we are all intelligent observers we will all figure out that the events were simultaneous in our rods-and-clocks reference frame." (graduate student)

An advanced undergraduate reached a similar conclusion after some clarification:

> I: Can you tell which order the eruptions occur in, in the spacecraft frame? Considered separately from the time of the light hit[ting] the spacecraft.
>
> S: I'm not really sure how to do that. It would seem to me that just logically, it doesn't matter, it would still go Rainier, Hood.
>
> I: You're speculating, if I can paraphrase you, that after the spacecraft [observer] made corrections for the fact that it took the Hood signal some time to get to him, after he made those corrections he would wind up concluding that Rainier went first?
>
> S: If he did everything right then he would have appropriately come up with the amount of time the signal would have traveled, the distance between the two mountains would probably also have been different to him, and so he'd have to account for that too, but once he made all of those accounting things then he would have to say that yes, they went up at the same time.
>
> I: Oh, at the same time. I thought you said Rainier went first.
>
> S: No, no, no, he would see Rainier go up first, but he would eventually after doing all of the math would agree with [two observers at rest on the ground], that they went up at the same time. (advanced undergraduate student)

These students appeared to believe that events simultaneous in the ground frame would be simultaneous according to an observer in any reference frame who made appropriate corrections for signal travel time. In the following example a graduate student compared the spacecraft observer to an observer on the ground under the spacecraft, at Mt. Rainier:

> "There is no real difference between the spacecraft and the [observer on the ground under the spacecraft]. Because I said the signals reach [that observer] at different times, but he can determine at which times the signals were emitted. …So I can do the same thing on the spacecraft: I might see the signals at different times but I can figure out that they happened at the same time." (graduate student)

Such responses indicated students' belief that simultaneity is absolute. Yet students, in their discussion of relativistic effects, appeared to have heard the idea of the relativity of simultaneity. How could these incompatible ideas coexist so openly? The resolution to this apparent contradiction came from in-depth probing.

> S: There are really two separate kinds of reference frames. There is the kind of reference frames with all those rods and clocks extending to infinity, like in [the textbook]. But in practice, nothing happens except right where you are. So really, your reference frame is something you carry around with you …
>
> I: There is this thing about simultaneity being relative, about events that are simultaneous in my reference frame not necessarily being simultaneous in another reference frame. Which kind of reference frame does that refer to?
>
> S: Relativity of simultaneity is this local thing. It's not the rods and clocks thing, because if we are intelligent, we correct for that. It's this thing that if I see them at different times, they occurred at different times in my reference frame. (graduate student)

Responses such as these seemed to indicate that the belief that simultaneity is absolute is deeply held.[27] Any *appearance* to the contrary is only that – a visual appearance due to differences in the reception of signals from the events. This belief was typical among students who asserted that the order of events in the spacecraft frame is the order in which signals arrive at the spacecraft.

> I: This thing about events that are simultaneous in one reference frame, not being simultaneous in another reference frame? Do you have a sense of where that comes from?
>
> S: Light has a finite speed, so it's going to take some time for the information to travel from point A to point B wherever the observer is. This is a pretty good example actually. One observer is right between the mountains and he sees them at the same time, the other observer is not and so he sees them at different times. (graduate student)

### 2. *Tendency to regard the Lorentz transformation for time as correcting for signal travel time*

In deriving the relativity of simultaneity, many instructors invoke the Lorentz transformation for time. We found that conceptual difficulties with reference frames and the time of an event can prevent students from interpreting appropriately the terms in the Lorentz transformations. In particular, many students appeared to believe that the Lorentz transformations constitute a correction for signal travel time:

> S: In the reference frame of the spacecraft, does Event 1 occur before, after, or at the same time as Event 2? … Before. Even though the spacecraft is traveling very fast, I would say that it's right next to Mt. Rainier so it's going to see Mt. Rainier go off, and even though it's traveling towards Mt. Hood and the light from Mt. Hood is traveling towards it, it will still take some amount of time for the information of Mt. Hood exploding to reach it.
>
> I: Okay, so it'll see Mt. Rainier first. In the spacecraft's reference frame, after he makes any corrections for signal travel time that might be appropriate –
>
> S: Are we including Lorentz transformations in that? (advanced undergraduate student)

The "desynchronization term" in the Lorentz transformation for time ($- v dx/c^2$) presented particular difficulty for students.[28] Several cited that term in support of the idea that the time of an event is influenced by the position of an observer relative to the event:

> "[This term] is the correction for the travel time of the light. The time I have to wait in my frame to *see* one event and then the next one." (graduate student)

The student above went on to express his confusion about the fact that the term is *proportional* to the speed of the spacecraft. He thought it should be inversely proportional since the travel time for the signal from Hood is reduced as the speed of the spacecraft in the ground frame is increased.

### 3. *Tendency to treat simultaneity as independent of relative motion*

Some students stated explicitly that relativity of simultaneity is not directly related to relative motion. Even graduate students expressed this idea. As in the preliminary investigation, some students appeared to believe that relative motion does play a role in the timing of events in the spacecraft frame – but only to the extent that it influences the reception of signals by the spacecraft.

*Commentary*

We have found that students often incorporate the relativity of simultaneity into their own conceptual framework in a way that allows them to continue to believe in absolute simultaneity. They do so by treating the time of an event as the instant at which that event is *seen* to occur by an observer and attributing the relativity of simultaneity to signal travel time. Such incorrect beliefs can insulate students from gaining an understanding of the relativity of simultaneity as a consequence of the invariance of the speed of light. Instructors and textbooks often admonish students to distinguish between the corrections of a finite signal travel time and the inevitability of the relativity of simultaneity.[29] Apparently, such admonitions are insufficient.

### C. Belief that every observer constitutes a distinct reference frame

The *Spacecraft* question discussed thus far had originally been designed to probe student understanding of simultaneity. The results suggested, however, that we needed to investigate more deeply their understanding of reference frames. We therefore developed other questions to probe student beliefs about simultaneity, reference frames, and the role of observers. Results from two of these, the *Explosions* question and the *Seismologist* question, are described below.

*Explosions question*

In the *Explosions* question, students are given the time interval between two non-simultaneous events in one frame and asked whether there is second frame in which the events are simultaneous. As in the first two versions of the *Spacecraft* question, no mention is made of a specific observer in the second frame. However, students often raised the issue of the observer location spontaneously.

The question has been given in written form to introductory students in sophomore modern physics courses and to students in a junior-level course on electricity in magnetism. In both cases, the question was administered on an examination after instruction in special relativity. We have also asked the *Explosions* question during interviews with 5 advanced undergraduate and 12 graduate students.

The percentage of correct responses on the written question was 40%. In the interviews, 3 of the 5 undergraduates and 7 of the 12 graduate students answered correctly. Many students attempted to solve the questions mathematically and it was often difficult to characterize their errors. However, the responses of students at all levels often reflected the following difficulty.

1. ***Tendency to treat observers at the same location as being in the same reference frame, independent of relative motion***

The following student correctly determined the relative speed of the frame in which the two explosions are simultaneous. [The student uses the rule developed in class that clocks in moving frames that follow (or "chase") other clocks read earlier times.] However, the student places an additional constraint on the solution by specifying a location for the observer such that the observer would see the explosions simultaneously.

"If we travel at a speed in which [one] side is the chasing side, it will be ahead by a certain time. Now if we set it to be $c\mathbf{d}t = 1200$ m ahead, *and we stand where the engineer is standing,* we'll see the explosions at the same time. So, you must be traveling at 0.4c, and must be at the point where the engineer is standing." (introductory student – italics added for emphasis)

The student seems to believe that the explosions are simultaneous only for one observer in the 'moving' frame (the observer who sees them simultaneously). Another introductory student answered similarly.

"There is no 'frame' that you will always see them [the two explosions] at the same instant, but there is a position. We can be [a certain number of] meters from the right [end] and see them at the same time." (introductory student)

The student seems to have interpreted the question "Is there a frame in which the events are simultaneous?" to mean "Is there an observer who sees the events at the same time?" This student apparently believes that a set of observers at rest relative to one another would not agree that the explosions are simultaneous since such observers would not all receive light from the two explosions at the same instant. The student was unable to apply the idea of a reference frame as a system for measuring the time of events.

*Seismologist question*

In the *Seismologist* question, students are asked about the relative ordering of two events for a seismologist and an assistant at rest relative to one another. The question was designed to probe whether students would incorrectly treat simultaneity as relative, even for two observers in the same reference frame.

The question has been administered to introductory and advanced students both as an examination question and as an ungraded written question. It has also been given during interviews to advanced undergraduates and graduate students. Relatively few students at any level answered correctly about the time order of events in the frame of the assistant. Only between 15% and 40% of the students answered correctly on the written questions. As shown in Table II, even advanced students had difficulty answering this question correctly. Many of the errors indicated the following tendency.

Table I. Results from the third *(location-specific)* version of the *Spacecraft* question. Students are told the eruptions are simultaneous for one observer and are given the location of an observer in a second frame moving relative to the first. They are asked for the order of the eruptions in the frame of the second observer. [Percentages have been rounded to the nearest 5%.]

| | Written question | | | | | | Interview task |
|---|---|---|---|---|---|---|---|
| | Non-physics students | | Introductory students (Honors calculus-based physics) | | Advanced undergraduates | | Advanced undergraduates and graduate students |
| | During instruction ($N = 40$) | After instruction ($N = 26$) | Before instruction ($N = 88$) | After instruction ($N = 79$) | Before instruction ($N = 48$) | After instruction ($N = 63$) | ($N = 17$) |
| | % (N) | % (N) | % (N) | % (N) | % (N) | % (N) | % |
| Correct answers (simultaneous eruptions) regardless of reasoning | 15% (6) | 10% (3) | 20% (19) | 30% (25) | 40% (20) | 40% (24) | 60% (10) |
| Rainier erupts first | 65% (25) | 75% (20) | 65% (57) | **60% (49)** | 55% (26) | 50% (33) | 40% (7) |
| Other (*e.g.,* Hood erupts first, student stated not enough information given) | 25% (9) | 10% (3) | 15% (12) | 5% (5) | 5% (2) | 10% (6) | 0 (0) |

### 2. *Tendency to treat observers at rest relative to one another as being in separate reference frames*

Essentially all of the students who answered the *Seismologist* question incorrectly stated that Mt. Rainier erupts first in the frame of the assistant. Some of these students were explicit about their interpretation of the term "reference frame."

> "Assuming the assistant is his reference frame, Rainier will erupt first because he will see its light first, and until he sees its light, effectively it hasn't erupted yet." (introductory student)

> "Reference frames, they're dependent on position, so that's why [the seismologist] and [the assistant] measure two different things. I think of them being in different reference frames simply because some people would say 'Well, if they're in the same reference frame everything should happen the same.' And it doesn't happen the same, because they measure two different times. That's why I am tempted to say although they're both at rest with respect to each other, they're in different reference frames." (graduate student)

> "... 'in the reference frame of the assistant' means you're sitting there and waiting for events to happen, and you record them when you see them, and that's when you mark them down so that's when they happen." (graduate student)

In the view of the student above, "the reference frame of the assistant" consists of a single observer (the assistant). Many students treated a reference frame as being local to the position of an observer.[30]

In one version of the *Seismologist* question, students are asked explicitly about both (1) the order in which the light signals from the eruptions reach the assistant and (2) the order of the eruption events in the reference frame of the assistant. In their responses, some students stated explicitly that the questions were identical.

"If by his reference frame, you mean, 'When did he see it?' it would be before."

Despite the fact that both observers are in the same frame, some students referred explicitly to the relativity of simultaneity or to the lack of synchronization of clocks.

"Mt. Rainier erupts first in the assistant's frame according to the relativity of simultaneity."

"According to the assistant Mt. Rainier erupted first because the two clocks (eruptions) are unsynchronized according to him and the light from Rainier is closer." (advanced undergraduate student)

The belief that each observer constitutes a separate reference frame was common and seems to be quite strongly held. The dialogue below provides an example.

S: The assistant is at Mt. Rainier, and Mt. Rainier erupts, and Mt. Hood erupts at the same time. But he can't see it the same time, because if you look at this mountain you always see in the past.

I: So if he made measurements of [times and distances], and he knew the speed of the signal and so on, what would he conclude about the eruption times? …

S: For the assistant Mt. Rainier would erupt before Mt. Hood. This is what I would say.

I: In the assistant's reference frame Rainier would erupt first?

S: Yeah, definitely.

I: And what if the assistant made measurements, and had assistants of his own, whatever measurements he needed to make to reach a conclusion about the timing of the eruptions? After all those measurements, the assistant would say, "In my reference frame, Mt. Rainier erupted first"?

S: Yes. (graduate student)

Unlike some of the introductory students, nearly all of the advanced students distinguished clearly between emission events (the eruptions) and reception events (the arrival of the light from an eruption at a particular observer). They also recognized that the travel time of light is relevant. However, nearly all of these students indicated a belief that such corrections were not appropriate for observers to make in attempting to determine the time of an event in their reference frames. During the interviews, many expressed the view that "reference frame" describes what an observer perceives at a particular location. One student went so far as to express the belief that the time ordering of events in an observer's frame depends on which signals that observer is able to detect:

"Within normal human ability to comprehend time, I would say that the eruptions are going to be at the same time. But if he's blind, he's going to hear Rainier for sure go off before Hood. And so he's going to say that Rainier went off before Hood because it's going to take much longer for the sound from Hood to get there." (graduate student)

The student quoted above was also asked to sketch pictures of the explosions as they happen in the seismologists' reference frame. She responded, "Which seismologist? The one at the base of the mountain, or the one in the middle?" At the interviewer's cautious response of "Both," she sketched two sets of pictures – one for the seismologist in the middle, and one for the assistant at Mt. Rainier. [See Fig. 4.] She explicitly indicated that the eruptions were simultaneous according to one observer but not the other.

Some students argued that observers at rest disagree on the time of an event even when the signal from that event explicitly contains the time of that event. The following exchange is illustrative.

I: Let's say … you are in the middle, with your fancy Rolex [watch], and I'm at Mt. Rainier, with my fancy Rolex. And you determine that at exactly noon, you took your vitamins. And suppose I wanted to figure out what time you took your vitamins.

S: If you're looking at me through a telescope you will look at my clock and it will say 12, but you will look at your clock and it will say 12 plus ?t.

I: So in my reference frame, did you take your vitamins at 12? Or at 12 plus ?t?

S: Twelve plus ?t. (advanced undergraduate student)

*Commentary*

When students remarked that the assistant can only know what happens at his own location, they were indicating that they had not understood the basic concept of a reference frame. They had not recognized that intelligent observers at rest with respect to one another can communicate about the spacetime coordinates of events at each location. Instead, these students seemed to think that each observer is restricted to the information that he or she can obtain directly.

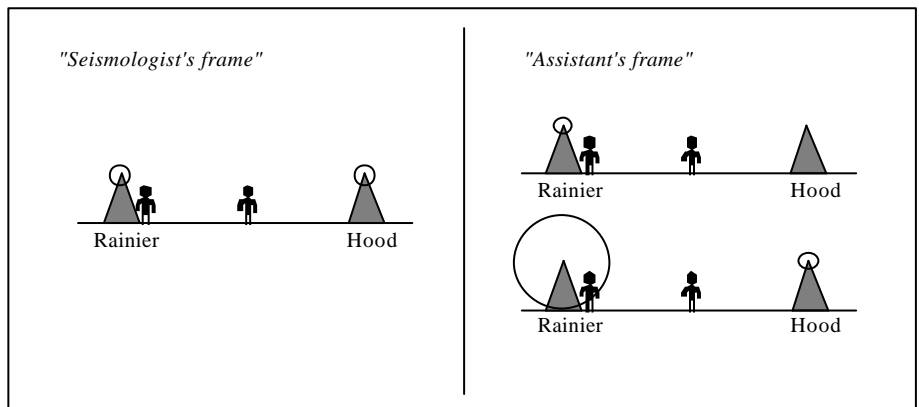

Figure 4: Student response to the *Seismologist* question. The student has indicated that the eruptions are simultaneous in the reference frame of the seismologist (in the middle), but that in the reference frame of the assistant (at Mt. Rainier), Mt. Rainier erupts first.

## VII. CONCLUSION

This investigation has identified widespread difficulties that students have with the definition of the time of an event and the role of intelligent observers. After instruction, more than 2/3 of physics undergraduates and 1/3 of graduate students in physics are unable to apply the construct of a reference frame in determining whether or not two events are simultaneous. Many students interpret the phrase "relativity of simultaneity" as implying that the simultaneity of events is determined by an observer on the basis of the *reception* of light signals. They often attribute the relativity of simultaneity to the difference in signal travel time for different observers. In this way, they reconcile statements of the relativity of simultaneity with a belief in absolute simultaneity and fail to confront the startling ideas of special relativity.

Experienced instructors know that students often have trouble relating measurements made by observers in different reference frames. It is not surprising that students, even at advanced levels, do not fully understand the implications of the invariance of the speed of light. What is surprising is that most students apparently fail to recognize even the basic issues that are being addressed. Students at all levels have significant difficulties with the ideas that form the foundations of the concept of a reference frame. In particular, many students do not think of a reference frame as a system of observers that determine the same time for any given event. Such difficulties appear to impede not only their understanding of the relativity of simultaneity, but also their ability to apply correctly the Lorentz transformations.

Special relativity offers instructors an opportunity to channel student interest in modern physics into a challenging intellectual experience. For most people, the implications of special relativity are in strong conflict with their intuition. For students to recognize the conflict and appreciate its resolution, they need to have a functional understanding of some very basic concepts. Formulating an appropriate measurement procedure for the time of an event involves recognizing the inherently local nature of measurement, applying a well-defined measurement procedure in a given reference frame, and understanding the relationship between measurements made by different observers. These ideas are crucial in contexts ranging from the rolling of a steel ball on a level track to the motion of objects in the vicinity of massive stars. This investigation documents prevalent modes of reasoning with these fundamental concepts as a first step toward making special relativity meaningful to students.

## ACKNOWLEDGMENTS

The investigation described in this paper has been a collaborative effort by many members of the Physics Education Group, present and past. We are extremely grateful to Lillian C. McDermott for her intellectual contributions in helping synthesize and clarify the issues addressed in the paper and her help in the writing. Bradley S. Ambrose and Andrew Boudreaux played significant roles in the initial stages of the research. Special thanks are also due to Paula R.L. Heron. Arnold Arons and Edwin F. Taylor gave us invaluable insights into the treatment of the relativity of simultaneity. Also deeply appreciated is the ongoing cooperation of the colleagues in whose physics classes the written problems have been administered, especially James Bardeen and E. Norval Fortson. We would also like to thank Tevian Dray and Greg Kilcup. The authors gratefully acknowledge the support of the National Science Foundation through Grants DUE 9354501 and DUE 9727648.

---

[1] For an extensive bibliography, see the relevant sections in L.C. McDermott and E.F. Redish, "Resource Letter: PER-1: Physics Education Research," Am. J. Phys. **67**, 755-767 (1999).

[2] For an article that illustrates how research can guide the development of curriculum, see L.C. McDermott, Millikan Award Lecture, "What

we teach and what is learned – Closing the gap," Am. J. Phys. **59**, 301-315 (1991).

[3] For examples of curriculum that have been developed through an iterative process of research, curriculum development, and instruction, see L.C. McDermott, P.S. Shaffer, and the Physics Education Group at the University of Washington, *Tutorials in Introductory Physics,* Preliminary Edition (Prentice Hall, Upper Saddle River, NJ, 1998); L.C. McDermott and the Physics Education Group at the University of Washington, *Physics by Inquiry* (Wiley, New York, NY, 1996), Vols. I and II.

[4] S. Panse, J. Ramadas, and A. Kumar, "Alternative conceptions in Galilean relativity: Frames of reference," Int. J. Sci. Educ. **16**, 63-82 (1994); J. Ramadas, S. Barve, and A. Kumar, "Alternative conceptions in Galilean relativity: Inertial and non-inertial observers," Int. J. Sci. Educ. **18**, 615-629 (1996).

[5] E. Saltiel and J.L. Malgrange, "'Spontaneous' ways of reasoning in elementary kinematics," Eur. J. Phys. **1**, 73-80 (1980).

[6] A. Villani and J.L.A. Pacca, "Students' spontaneous ideas about the speed of light," Int. J. Sci. Educ. **9**, 55-66 (1987) and "Spontaneous reasoning of graduate students," Int. J. Sci. Educ. **12**, 589-600 (1990).

[7] P.W. Hewson, "A case study of conceptual change in special relativity: The influence of prior knowledge in learning," Eur. J. Sci. Educ. **4**, 61-76 (1982).

[8] G.J. Posner, K.A. Strike, P.W. Hewson, and W.A. Gertzog, "Accommodation of a scientific conception: Toward a theory of conceptual change," Sci. Ed. **66**, 211-227 (1982).

[9] T.E. O'Brien-Pride, "An investigation of student difficulties with two dimensions, two-body systems, and relativity in introductory mechanics," Ph.D. dissertation, Department of Physics, University of Washington, 1997 (unpublished).

[10] That these are concepts that require definition is itself a new idea to many students. For insightful discussions of these definitions and the pedagogical concerns that they raise, see, for instance, P.W. Bridgman, *A sophisticate's primer of relativity* (Wesleyan University Press, Middletown, CT, 1962) and A.B. Arons, *A guide to introductory physics teaching* (Wiley, New York, NY, 1990).

[11] The definition of a global coordinate system breaks down in non-inertial frames and in general relativity. The restriction of an inertial frame to a finite extent in both space and time is a refinement not usually encountered in courses in special relativity.

[12] One method of synchronizing clocks in special relativity includes sending the reading on one clock to a clock at another location by means of some signal. The second clock is synchronized with the first by setting it to read the time sent from the first clock plus the signal travel time. The use of light signals for the synchronization of clocks is customary but not necessary. See, for instance, the first book in Ref. 10.

[13] Although it is possible to define the time of an event as the time at which an observer *sees* the event, the time then depends on observer location. Einstein, for instance, considered and rejected such a definition. See A. Einstein, "On the electrodynamics of moving bodies," in *The principle of relativity: A collection of original memoirs on the special and general theory of relativity* (Dover, New York, NY, 1952).

[14] Some authors define the term "observer" to indicate the full set of measuring devices and procedures that comprise a reference frame. For an example of this approach, see E.F. Taylor and J.A. Wheeler, *Spacetime Physics* (W.H. Freeman, New York, NY, 1992), pp. 39.

[15] The process by which "time is spread over space" is described meticulously in the first book in Ref. 10.

[16] For a description of the course for high school teachers, see L.C. McDermott, "A perspective on teacher preparation in physics and other sciences: The need for special courses for teachers," Am. J. Phys. **58**, 734-742 (1990).

[17] The results from the various classes at the University of Washington were consistent within statistical fluctuations. The results from the other universities were within the same range. For the purposes of this article,

[17] the results from corresponding classes have been combined.

[18] The problems are "ungraded" in the sense that they are not graded for correctness. Rather, students receive credit for giving responses that reflect an attempt to answer the questions. Student performance on the examination questions and the ungraded problems was very similar. This is consistent with the results of other investigations conducted by our group, in which we have found that students take the ungraded questions seriously.

[19] Each research question was posed in at least two ways, with different context and slight changes in wording. In general, we found that such changes had little effect on student performance. Therefore, in this paper we present only a representative description or example of each question.

[20] The fact that the curved surface of a gravitating, rotating Earth is not an inertial frame did not elicit student concern. In one version of the question, students were told to neglect non-inertial effects. In another, the context was set in deep space. Neither statement seemed to change student responses.

[21] In keeping with standard practice in one-dimensional problems on special relativity, students were told in all questions that all motions were to be considered as occurring along a single line in space. No student seemed to have had difficulty neglecting the vertical dimension.

[22] The prototype of such a qualitative analysis is that of the classic train paradox often used to develop the relativity of simultaneity. See, for instance, P.A. Tipler and R.A. Llewellyn, *Modern Physics* (W.H. Freeman, New York, NY, 1999).

[23] We use the notation *d* instead of ?*t, etc.,* to try to minimize confusion between *the difference between two quantities* and *a change in a quantity*. We are grateful to Eric Mazur for a discussion on this point.

[24] Since the events are separated by a spacelike interval ($c^2 \delta t^2 - \delta x^2 < 0$), students can predict the *existence* of a frame in which the events are simultaneous without use of the Lorentz transformations.

[25] We have evidence that the students in this class performed as well as they did on the directed version of the *Spacecraft* question as a result of special instruction that they had received. On the basis of the research described in this paper, we have been developing instructional materials to address the specific difficulties that we have identified. This class had used preliminary versions of these instructional materials. As is demonstrated in section VI, other classes after standard instruction did not do as well on similar questions.

[26] This finding is consistent with our experience that the study of advanced material does not necessarily deepen conceptual understanding. See, for example, S. Vokos, P.S. Shaffer, B.S. Ambrose, and L.C. McDermott, "Student understanding of the wave nature of matter: Diffraction and interference of particles," Phys. Educ. Res., Am. J. Phys. Suppl. **68**, S42-S51 (July 2000); B.S. Ambrose, P.S. Shaffer, R.N. Steinberg, and L.C. McDermott, "An investigation of student understanding of single-slit diffraction and double-slit interference," Am. J. Phys. **67**, 146-155 (1999); K. Wosilait, P.R.L. Heron, P.S. Shaffer, and L.C. McDermott, "Development of a research-based tutorial on light and shadow," *ibid.* **66**, 906-913 (1999); T. O'Brien Pride, S. Vokos, and L.C. McDermott, "The challenge of matching learning assessments to teaching goals: An example from the work-energy and impulse-momentum theorems," *ibid.* **66**, 147-157 (1998); L.C. McDermott, P.S. Shaffer, and M.D. Somers, "Research as a guide for teaching introductory mechanics: An illustration in the context of the Atwood's machine," *ibid.* **62**, 46-55 (1994).

[27] The belief in absolute simultaneity may be related to the strong belief of students in a preferred reference frame, which has been documented in the context of Galilean relativity. See Refs. 5, 6, and 9.

[28] This term is related to the amount of time that two specific synchronized clocks in the spacecraft frame are measured to be out of synchronization by observers in the ground frame.

[29] See, for instance, D. Griffiths, *Introduction to Electrodynamics* (Prentice Hall, Upper Saddle

River, NJ 1989), p. 452. This widely used text states explicitly that the relativity of simultaneity is "a genuine discrepancy between measurements made by competent observers in relative motion, not a simple mistake arising from a failure to account for the travel time of light signals."

[30] The belief that each observer constitutes a distinct reference frame is similar to the belief documented in Galilean contexts in Ref. 4. The authors describe a tendency of students to treat the extent of an observer's reference frame as limited to the physical object on which the observer is located (*e.g.,* the deck of a boat).